\documentclass[twocolumn,times]{aastex631}
\usepackage{amsmath,bm}
\usepackage{xcolor}
\usepackage[english]{babel}
\usepackage[latin1]{inputenc} 
\usepackage{enumitem}

\newcommand{\msun}{M_{\sun}}
\newcommand{\msunh}{h^{-1}M_{\sun}}

\newcommand{\hi}{H{~\sc i}}

\shorttitle{Central Galaxy Star Formation and Quenching}
\shortauthors{H. Guo et al.}

\begin{document}

\title{Cold Gas Reservoirs of Low- and High-mass Central Galaxies Differ in Response to Active Galactic Nucleus Feedback} 

	\author[0000-0003-4936-8247]{Hong Guo}
	\altaffiliation{Corresponding Author}
	\affiliation{Shanghai Astronomical Observatory, Chinese Academy of Sciences, Shanghai 200030, China; guohong@shao.ac.cn}
	
	\author[0000-0002-5434-4904]{Michael G. Jones}
	\altaffiliation{Corresponding Author}	
	\affiliation{Steward Observatory, University of Arizona, 933 North Cherry Avenue, Rm. N204, Tucson, AZ 85721-0065, USA; jonesmg@arizona.edu}
	
	\author[0000-0002-6593-8820]{Jing Wang}
	\altaffiliation{Corresponding Author}	
	\affiliation{Kavli Institute for Astronomy and Astrophysics, Peking University, Beijing 100871, China; jwang\_astro@pku.edu.cn}

\begin{abstract}
The growth of supermassive black holes, especially the associated state of active galactic nuclei (AGNs), is generally believed to be the key step in regulating star formation in massive galaxies. As the fuel of star formation, the cold gas reservoir is a direct probe of the effect of AGN feedback on their host galaxies. However, in observation, no clear connection has been found between AGN activity and the cold gas mass. In this paper, we find observational signals of significant depletion of the total neutral hydrogen gas reservoir in optically-selected type-2 AGN host central galaxies of stellar mass $10^{9}$--$10^{10}M_\odot$. The effect of AGN feedback on the cold gas reservoir is stronger for higher star formation rates and higher AGN luminosity. But it becomes much weaker above this mass range, consistent with previous findings focusing on massive galaxies. Our result suggests that low-mass and gas-rich AGN host central galaxies would first form dense cores before AGN feedback is triggered, removing their neutral hydrogen gas. More massive central galaxies may undergo a significantly different formation scenario by gradually building up dense cores with less effective and recurrent AGN feedback.
\end{abstract}

\section{Introduction}
In current galaxy formation theories, feedback from an active galactic nucleus (AGN) is regarded as one of the most effective channels of shutting down star formation, through the depletion and heating of cold gas within and surrounding a galaxy \citep[see e.g.,][for reviews]{Heckman2014,Schreiber2020}. While AGN-driven outflows and winds have been known for decades \citep[e.g.,][]{Fabian2012}, it is still controversial whether these can sufficiently reduce the overall gas reservoir to trigger quenching of star formation \citep{Harrison2017}. 

Since the quenching of satellite galaxies is found to be mostly driven by the halo environmental effect of cold gas depletion \citep[e.g.,][]{Wetzel2013,Tal2014,Jaffe2015,Brown2017,Stevens2019}, we only focus on central galaxies in this study. In our previous work \citep[][hereafter G21]{Guo2021}, we find that the star formation activity of central galaxies in the local universe is directly regulated by the available atomic neutral hydrogen (\hi) reservoir . It is therefore essential to quantify the effect of AGN feedback on the \hi\ gas mass in order to understand its role in the quenching of star formation. 

However, previous observational efforts to measure the cold gas mass in AGN host galaxies revealed no strong dependence on AGN luminosity for both \hi\ mass \citep{Fabello2011,Gereb2015,Ellison2019} and molecular gas mass \citep[e.g.,][]{Shangguan2020}, challenging current galaxy formation models.

To better understand the controversy, here we apply an established \hi\ spectra stacking technique (G21) to a large statistical sample of  11240 star-forming galaxies (SFGs) and 6368 type-2 AGN host galaxies from Sloan Digital Sky Survey \citep[SDSS;][]{SDSSDR7} DR7. Different to previous studies, we compare their \hi\ reservoir at the same stellar mass ($M_\ast$) and star formation rate (SFR) bins, i.e. $M_{\rm HI}({\rm SFR}|M_\ast)$, rather than comparing $M_{\rm HI}(M_\ast)$ for AGN hosts and non-AGN control galaxies as in \cite{Ellison2019}. As will be shown below, such a division is essential to isolate the influence of AGN hosts from their compound dependence on both $M_\ast$ and SFR. We further extend the work of \cite{Ellison2019} by including the effect of central stellar surface density within 1~kpc (denoted as $\Sigma_1$), which is found to be tightly correlated with $M_{\rm HI}$ in G21.    

The structure of this paper is as follows. We describe the galaxy samples and stacking method in \S\ref{sec:data}. We present the results in \S\ref{sec:results}. We summarize and discuss the results in \S\ref{sec:discussion}. Throughout the paper, the halo mass is in units of $\msunh$, while the stellar and \hi\ masses are in units of $\msun$.

\begin{figure*}
	\centering\includegraphics[width=0.9\linewidth]{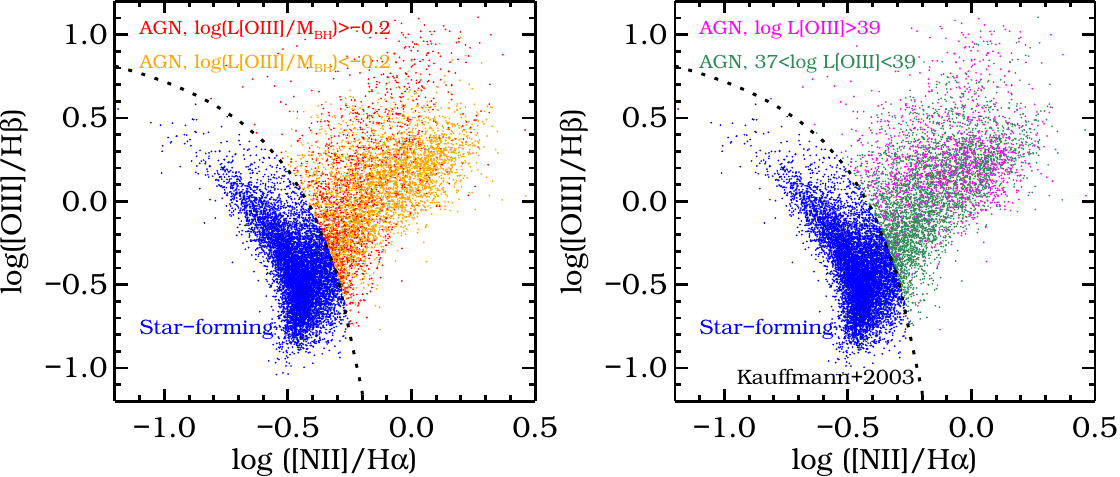}
	\caption{Sample definition. The SFGs and AGN hosts (represented by dots of different colors) are classified with the demarcation line \citep{Kauffmann2003} (dotted line). The AGNs are divided into subsamples using the Eddington parameter (ratio between [O{\sc iii}] luminosity $L_{\rm [O{\sc III}]}$ and black hole mass $M_{\rm BH}$, left panel) and $L_{\rm [O{\sc III}]}$ (right panel). The characteristic value of $\log(L_{\rm [O{\sc III}]}/M_{\rm BH})\sim-0.2$ roughly corresponds to 1\% of the Eddington ratio \citep{Kauffmann2009}, separating the AGNs into radiative and jet modes \citep{Heckman2014}.}
	\label{fig:sample}
\end{figure*}
\section{Data and Method}\label{sec:data}
The galaxy sample in this work is the same as in G21. The galaxies used for the \hi\ spectra stacking are selected from the overlap regions between the optical data of SDSS DR7 and \hi\ 21~cm data of the Arecibo Fast Legacy ALFA Survey \citep[ALFALFA;][]{Giovanelli2005,Haynes2018} 100\% complete catalog in the redshift range of $0.0025<z<0.06$. The central galaxies are identified with a galaxy group catalog based on SDSS \citep{Lim2017}. 

We adopt the galaxy stellar mass and SFR measurements from the GSWLC-2 catalog \citep{Salim2018}, where the UV/optical spectral energy distribution fitting was applied to have reliable SFRs for both SFGs and AGN hosts \citep{Salim2016}. We also use the measurements of $\Sigma_1$ (in units of $M_\odot/\rm{kpc}^2$) as in G21, which is obtained from the product of the total light within 1~kpc and the $i$-band mass-to-light ratio. 

The SDSS galaxy emission line fluxes and stellar velocity dispersion measurements are obtained from the MPA-JHU DR7 release\footnote{https://wwwmpa.mpa-garching.mpg.de/SDSS/DR7/}. Based on the BPT diagram \citep{Baldwin1981,Veilleux1987}, we select type-2 AGNs from galaxies with S/N$>3$ in all four emission lines of [O{\sc iii}]$\lambda$5007, [N{\sc ii}]$\lambda$6583, H$\alpha$ and H$\beta$, following the demarcation line \citep{Kauffmann2003}, $\log([{\rm O III}]/{\rm H}\beta)>0.61/[\log([{\rm N II}]/{\rm H}\alpha])-0.05]+1.3$.

Taking advantage of the large statistical sample, we are able to separate the AGN hosts into different $M_\ast$ and SFR bins. As shown in Figure~\ref{fig:sample}, the subsamples are further divided according to the [O{\sc iii}] luminosity ($L_{\rm [O{\sc III}]}$, in units of $\rm{erg/s}$, right panel) and the Eddington parameter ($L_{\rm [O{\sc III}]}/M_{\rm BH}$, a proxy of Eddington ratio suggested in \citealt{Heckman2004}, left panel). The [O{\sc iii}] luminosity has been corrected for dust extinction using the Balmer decrement by assuming a dust attenuation law of $\tau_{\lambda}\propto\lambda^{-0.7}$ \citep{Charlot2000}. The black hole mass $M_{\rm BH}$ is estimated using the stellar velocity dispersion--black hole mass relation \citep{Tremaine2002}. The characteristic value of $\log(L_{\rm [O{\sc III}]}/M_{\rm BH})\sim-0.2$ roughly corresponds to 1\% of the Eddington ratio \citep{Kauffmann2009}, separating the AGNs into radiative and jet modes \citep{Heckman2014}. 

For the \hi\ mass measurements, we follow our previous \hi\ spectra stacking technique (G21), an improvement of the methods of previous ALFALFA-based stacking works \citep{Fabello2011,Guo2020}. Adaptive aperture sizes were applied for galaxies with different stellar masses, as $\log(D_\mathrm{aper}/\mathrm{kpc}) = 0.130 \log(M_\ast/\mathrm{M_\odot}) + 0.635$, with a lower limit of $4^\prime$ to account for the resolution of Arecibo telescope at 21~cm. With the adaptive aperture sizes, the effect of confusion caused by the Arecibo beam size is minimized. We estimate that the typical correction of the stacked \hi\ mass is found to be only around 0.05~dex, which would not significantly affect our results. The measurement errors of the stacked \hi\ masses were estimated from the statistical uncertainties in the stacked spectra, which provides a comparable estimate to the bootstrapping method. These are typically quite small with a large number of stacked spectra. We refer the readers to G21 for more details. 

After the sample selection, our galaxy catalog consists of 8118 AGN hosts and 13672 SFGs. The galaxy \hi\ spectra with excessive noise are discarded to achieve the best S/N, since most of the noise is caused by the radio frequency interference (RFI) that is not related to the \hi\ signal  \citep{Guo2020}. Finally, there are 6368 AGN hosts and 11240 SFGs used in the direct spectra stacking. Moreover, only 1607 (25\%) AGN hosts and 5455 (49\%) SFGs in the stacked galaxies have available individual \hi\ mass measurements from ALFALFA. The \hi\ detection rate for the AGN hosts increases from 14\% in $10^{9.5}$--$10^{10}\msun$ to 27\% in $10^{10.5}$--$10^{11}\msun$, while that for SFGs is roughly constant around 48\%, which is indicative of the decreased \hi\ reservoir in AGN hosts compared to the star-forming counterparts at the low-mass end. It also makes the \hi\ stacking method important for the quantitative studies of the cold gas reservoir. 

\begin{figure*}
	\centering\includegraphics[width=0.9\linewidth]{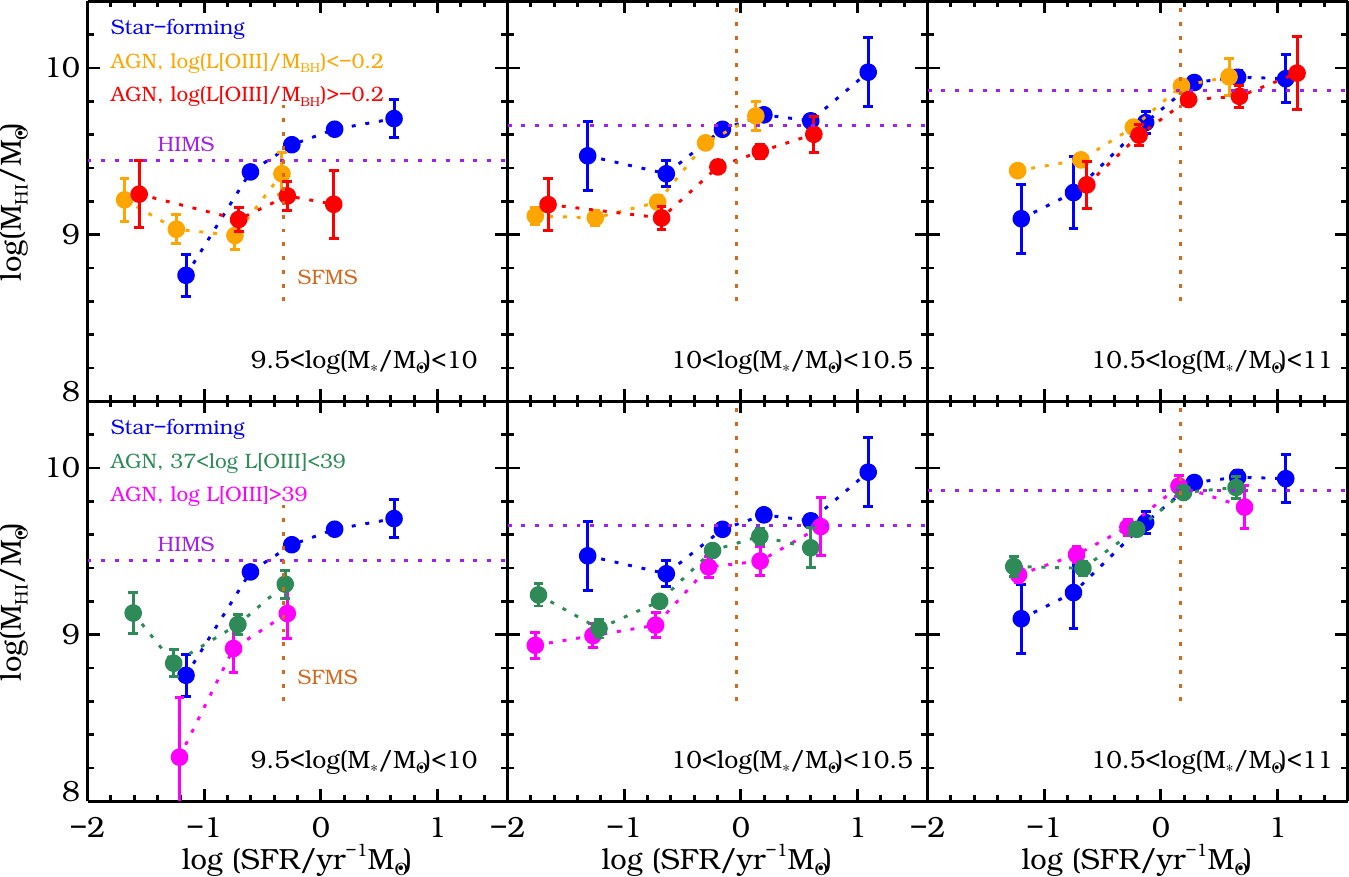}
	\caption{Stacked \hi\ measurements in $M_\ast$ and SFR bins. The stacked \hi\ masses are measured for galaxies in three stellar mass bins from $10^{9.5}M_\odot$ to $10^{11}M_\odot$ and 7 SFR bins from $10^{-2}M_\odot/\mathrm{yr}$ to $10^{1.5}M_\odot/\mathrm{yr}$. The measurements for AGNs are shown for the divisions in Eddington parameter (upper panels) and [O{\sc iii}] luminosity (lower panels), respectively. The symbol colors are the same as in Fig.~\ref{fig:sample}. The SFR measurements shown are the median values in each bin. The vertical lines are the locations of the star formation main sequence in each mass bin (Equation~2 of G21), and the horizontal lines represent the corresponding \hi\ main sequence ($\log M_{\rm HIMS}=0.42\log M_\ast+5.35$) for star-forming galaxies (G21). Some of the error bars for the \hi\ masses are invisible due to their typically small values (less than $0.05$) using stacked measurements.}
	\label{fig:hisfr}
\end{figure*}
\section{Results} \label{sec:results}
As shown in Fig.~\ref{fig:hisfr}, we split the galaxy sample into three $M_\ast$ bins of $10^{9.5}$--$10^{10}M_\odot$, $10^{10}$--$10^{10.5}M_\odot$ and $10^{10.5}$--$10^{11}M_\odot$ and measure the stacked \hi\ masses for SFGs and AGN hosts in the same SFR bins, which ensures that the comparison is not affected by the different SFR distributions of the two populations \citep{Ellison2019}. The star-formation main sequence (SFMS, G21) and the corresponding \hi\ main sequence (HIMS, defined for star-forming galaxies in G21, $\log M_{\rm HIMS}=0.42\log M_\ast+5.35$) are represented by the vertical and horizontal dotted lines, respectively.

There is a clear trend that lower-mass AGN hosts have significantly depleted \hi\ reservoirs compared to their star-forming counterparts (left panel). This effect is even stronger for higher SFRs, more luminous AGNs and also those with higher accretion rates (i.e. larger $L_{\rm [O{\sc III}]}/M_{\rm BH}$). In the lowest mass sample, the \hi \ masses of AGN hosts are smaller than their star-forming counterparts by up to $\sim$0.5~dex. This offset is not affected by the bin size in $M_\ast$ (0.5~dex), as AGN hosts are typically 0.1~dex more massive than the SFGs in each mass bin. That means the difference between AGN hosts and SFGs is even larger in terms of the \hi\ gas fraction ($f_{
\rm HI}\equiv M_{\rm HI}/M_\ast$). It is also not caused by the bin size of SFR (0.5~dex), as the SFR values shown in each bin are the median measurements. We show in G21 that for galaxies at a given $M_\ast$, the average relation between SFR and $M_{\rm HI}$ is around SFR$\propto M_{\rm HI}^{2.75}$, i.e. to account for the 0.5~dex decrease in $M_{\rm HI}$, the SFR needs to be decreased by 1.38~dex, much larger than the bin size effect. 

However, for massive galaxies with $\log(M_\ast/M_\odot)>10.5$, the AGN hosts and SFGs have very similar \hi\ masses, consistent with previous studies \citep{Fabello2011,Gereb2015,Ellison2019}. This shows that the global star formation law between SFR and $M_{\rm HI}$ for the most massive galaxies is not affected by the AGN activity, regardless of the Eddington parameter or [O{\sc iii}] luminosity.

Since both the measurements of AGN activity and \hi\ masses are instantaneous, as opposed to the time-averaged SFR, these thus provide observational signature of cold gas depletion in low-mass galaxies hosting optical type-2 AGNs. As quenched galaxies at $z\sim0$ on average have \hi\ masses $\sim$0.6~dex lower than the SFGs at all stellar masses (G21), the negative AGN feedback would thus act as the dominant mechanism of star-formation quenching in the stellar mass range $10^{9.5}$--$10^{10}M_\odot$, while its contribution seems to gradually decrease as stellar mass increases.

Previously, no significant difference was found in \hi\ fractions of AGN and non-AGN control galaxies for the mass range probed in this work \citep{Ellison2019}. This discrepancy is partially caused by the fact that here we exclude those galaxies with low S/N in the four emission lines of the BPT diagram, which could otherwise lead to galaxies being mis-classified. These galaxies with weak emission lines were included in the non-AGN control samples in the previous works \citep{Fabello2011,Ellison2019}. Another important difference is that the comparisons between AGN hosts and non-AGNs in the previous works were made with the stacked signals of $f_{\rm HI}(M_\ast)$, unlike in our case of $M_{\rm HI}({\rm SFR}|M_\ast)$. The essential dependence of $M_{\rm HI}$ on SFR is marginalized in $f_{\rm HI}(M_\ast)$. Although in each mass bin, the AGN population is dominated by those galaxies with lower SFRs compared to the SFGs, the effect of AGN feedback is, however, stronger at higher SFRs. Therefore, constructing the non-AGN control samples by matching both the stellar mass and SFR of AGN hosts (as in the previous work) would significantly weaken the signal of gas depletion, as the apparent differences shown in $M_{\rm HI}({\rm SFR}|M_\ast)$ at the high SFRs will be largely down-weighted by the SFR distribution peaked at the low SFRs in the integrated $M_{\rm HI}(M_\ast)$. This is further elaborated in the Appendix~\ref{sec:comparison}. 

\begin{figure*}
	\centering\includegraphics[width=1\linewidth]{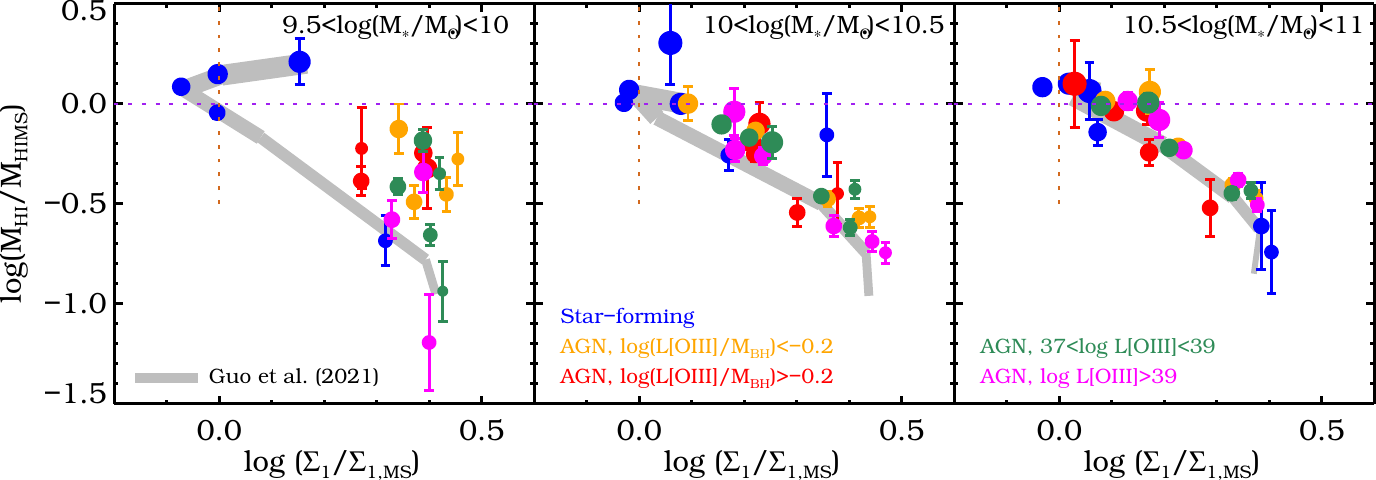}
	\caption{Relation between \hi\ mass and $\Sigma_1$ parameter. Both the \hi\ and $\Sigma_1$ measurements are scaled by their corresponding values from galaxies in the SFMS. The $\Sigma_1$ measurements shown are the median values in each $M_\ast$ and SFR bin. The symbols are the same as in Fig.~\ref{fig:hisfr}, with the symbol sizes representing the values of SFRs, with larger sizes for higher SFRs. The measurements of all galaxies including those unclassified galaxies in different $M_\ast$ and SFR bins from G21 are shown as the gray bands, with thicker lines indicating higher SFRs. The positions of AGN hosts in the mass range of $10^{9.5}$--$10^{10}M_\odot$ are well separated from the SFGs, with much higher $\Sigma_1$ and significantly decreased $M_{\rm HI}$. In more massive galaxies, they are approaching each other to form a tight $M_{\rm HI}$--$\Sigma_1$ relation.    
}
	\label{fig:hims}
\end{figure*}
The \hi\ mass of a galaxy has been found to be tightly correlated with the central stellar surface density within 1~kpc ($\Sigma_1$) in the quenching phase (G21), with $M_{\rm HI}\propto\Sigma_1^{-2}$. To further investigate the effect of AGN on the quenching process, we show in Fig.~\ref{fig:hims} the relation between $M_{\rm HI}$ and $\Sigma_1$ for the SFGs and AGN hosts in different $M_\ast$ and SFR bins, where $M_{\rm HI}$ and $\Sigma_1$ are normalized by the main sequence values of $M_{\rm HIMS}$ and $\Sigma_{\rm 1,MS}$ ($\log\Sigma_{\rm 1,MS}=0.81\log M_\ast+0.607$ as in G21), respectively. The measurements of all galaxies (including unclassified galaxies) from G21 in each bin are also shown (gray bands). 

In the mass range of $10^{9.5}$--$10^{10}M_\odot$, the overall relation is dominated by the SFGs due to their larger sample size. It is remarkable that all AGN hosts, irrespective of the  Eddington parameter or [O{\sc iii}] luminosity, have $\Sigma_1$ measurements well above $\Sigma_{\rm 1,MS}$ by at least 0.25~dex, even for AGN hosts with high SFRs. As the SFR values are indicated by the symbol sizes, we find that the concentrated distribution of the AGN hosts in the figure is not caused by the selection effects of their SFRs. In fact, these AGN hosts have on average 0.4~dex higher $\Sigma_1$ than the corresponding SFGs in the same SFR bins. The AGN activity is therefore associated with the increase of $\Sigma_1$ and decrease of $M_{\rm HI}$. 

As the correlation between $M_{\rm HI}$ and $\Sigma_1$ becomes much weaker for the AGN hosts at various SFRs, it may indicate that AGN luminosity is the main driving force of cold gas depletion in these low-mass galaxies after they form the dense cores through the compaction processes \citep{Dekel2014,Zolotov2015}. Their SFRs will soon decrease due to the loss of the \hi\ gas, leading to a quenched state before the next episode of star formation (likely indicated by the lowest blue dot shown in the left panel of Fig.~\ref{fig:hims}).

We further emphasize that $\Sigma_1$ is not equivalent to the global morphology of galaxies. As demonstrated in G21, the $M_{\rm HI}$-SFR relation does not depend on the galaxy morphology. The reduced $M_{\rm HI}$ in AGN hosts is then not due to their morphology changes with respect to the SFGs, as we also compare the gas reservoir at the same SFR bins. In fact, as discussed in \cite{Chen2020}, $\Sigma_1$ is increasing with $M_{\rm BH}$. The dependence of $M_{\rm HI}$ on $\Sigma_1$ thus reflects the gas stripping caused by the black hole growth, which will be explored in our upcoming work.

For more massive galaxies above $10^{10}M_\odot$, the distribution of AGN hosts in the diagram moves closer to the SFGs and shows a strong dependence of $M_{\rm HI}$ on $\Sigma_1$. They are almost indistinguishable from each other for $M_\ast>10^{10.5}M_\odot$ and form a tight $M_{\rm HI}$--$\Sigma_1$ relation, consistent with the case in Fig.~\ref{fig:hisfr} for the $M_{\rm HI}$--SFR relation. Such a similarity in the \hi\ content of AGN hosts and SFGs does not necessarily mean that AGN feedback is not effective in these massive galaxies. It has been proposed that the instantaneous mass accretion rate traced by the AGN luminosity could vary significantly during a typical star formation episode \citep{Novak2011,Hickox2014,Harrison2017}, making the difference between SFGs and AGN-dominated galaxies less apparent in the global \hi\ reservoir. One other possibility is that the AGN-driven outflows and jets only directly affect the inner regions of the host galaxies within a few kpc \citep{Karouzos2016,Ellison2021,Shi2021}, while the global cold gas reservoir roughly remains the same. 

It is important to note that the typical \hi\ disc diameter scales with stellar mass as $D_{\rm HI}\propto M_\ast^{0.5}$. \citep{Wang2016} For example, $D_{\rm HI}$ will decrease from about 60~kpc to 30~kpc when $M_\ast$ decreases from $10^{10}M_\odot$ to $10^{9.5}M_\odot$. It potentially makes the effect of AGN luminosity on the \hi\ gas more significant for lower-mass galaxies, given that there is only weak dependence of $L_{\rm [O{\sc III}]}$ on stellar mass \citep{Kauffmann2003}, with the average $L_{\rm [O{\sc III}]}$ remaining roughly constant at $\sim10^{38.3}\,{\rm erg/s}$ for $M_\ast<10^{10.5}\msun$.  

The higher gas fractions of low-mass galaxies supply more fuel for the central supermassive black holes to reach the high-accretion state, while the shallower gravitational potential wells mean that gas is more loosely bound to the galaxies. Combined with the fact that in massive galaxies the AGNs are transitioning to the jet mode (or maintenance mode), the cumulative energy release by the recurrent low-luminosity radio-AGN activity will prevent the hot halo gas from further cooling and allow for an efficient self-regulating AGN cycle, which keeps the global \hi\ reservoir at a similar level to the SFGs \citep{Heckman2014}.   

\section{Conclusions and Discussions}\label{sec:discussion}
In this paper, we compare the stacked \hi\ masses for SFGs and type-2 AGN hosts at the same $M_\ast$ and SFR intervals in the redshift range $0.0025<z<0.06$. We find that AGN hosts have systematically smaller \hi\ reservoir than their star-forming counterparts (by a maximal amount of $\sim0.5$~dex) with the same $M_\ast$ and SFR in the stellar mass range $10^{9.5}$--$10^{10}M_\odot$. This effect is even stronger for AGN hosts with higher SFRs, [O{\sc iii}] luminosity and Eddington ratios, providing observational support for the effect of AGN luminosity on the gas depletion. For more massive galaxies, the difference in $M_{\rm HI}$ is, however, significantly smaller, consistent with previous literature probing the same mass ranges.

Our detection of reduced \hi\ reservoir in low-mass AGN hosts suggests that the instantaneous AGN feedback is likely more effective in gas-rich host galaxies. These AGN hosts first grow dense cores, represented by their higher $\Sigma_1$ values than SFGs of the same mass. This is the so-called ``blue-nugget'' phase \citep{Dekel2014}. As proposed by theoretical models \citep{Barro2013,Cui2021}, the dissipative gas inflow would then lead to rapid central black hole growth, thus triggering AGN feedback and depleting the surrounding cold gas \citep{Chen2020}. The prerequisite for such an effect to be observable is that the AGN feedback is so violent that the global cold gas content, especially those distributed in the outer parts of galaxies, can be significantly depleted within the timescale of an AGN activity cycle (typically much less than 100~Myr). This probably requires the AGN hosts to be small. Therefore, the different behaviors of $M_{\rm HI}$--SFR as the masses of AGN hosts increase, may reflect the change from instantaneous to integrated AGN feedback effects, along with the changes in AGN modes. 

For an order-of-magnitude estimation, the gravitational potential energy of $10^9M_\odot$ \hi\ gas (with $D_{\rm HI}\sim 30\,$kpc) exerted by all the stars in a $10^{9.5}M_\odot$ galaxy is roughly $10^{55}$~erg, which is similar to the energy required to fully ionize the gas. The gravitational potential energy from the dark matter is ignored, as the depleted gas does not necessarily leave the host halo. If we take the lower threshold of the AGN luminosity $L_{\rm [O{\sc III}]}=10^{39}\,{\rm erg/s}$, along with a factor of 600 for the bolometric correction \citep{Kauffmann2009},  and 10\% of energy release to the surrounding gas, the cumulative energy released within 5~Myrs also reaches $10^{55}$~erg. Although such an estimation is quite uncertain without detailed models and it is not really required to expel all the gas, it at least suggests that the feedback from AGNs to drive the gas depletion is quantitatively feasible, as implemented in the modern hydrodynamical simulations \citep{Weinberger2018}.

We emphasize that the causal link between \hi\ depletion and AGN luminosity cannot be directly inferred from our current measurements. But compared to other mechanisms, the AGN feedback is still the most likely cause. The stellar feedback is generally found to be dominant for galaxies with $M_\ast<10^9\msun$ and the energy injected by stellar winds is typically proportional to SFR \citep{Weinberger2018}. Our comparisons of $M_{\rm HI}({\rm SFR}|M_\ast)$ are thus made for galaxies with similar levels of stellar feedback, further highlighting the influence of AGN luminosity. 

The AGN luminosity might also be effective in even lower mass galaxies, below $10^{9.5}M_\odot$, as inferred from a possible deficit gas fraction in AGN hosts at these masses \citep{Bradford2018,Ellison2019}. Due to the low AGN fraction, when extending our sample to the range of $10^{9}$--$10^{9.5}M_\odot$, we can only obtain \hi\ measurements for AGN hosts in two $\log({\rm SFR}/{\rm yr}^{-1}M_\odot)$ bins of $[-1.5,-1]$ and $[-1,-0.5]$, which are just around the SFMS at this mass range. The corresponding $\log(M_{\rm HI}/M_\odot)$ measurements are $7.63\pm0.49$ ($8.99\pm0.03$) and $8.10\pm0.20$ ($9.31\pm0.02$) for all AGNs (SFGs), respectively. In terms of the $\log(\Sigma_1/\Sigma_{\rm 1,MS})$ measurements, they are 0.36 (-0.01) and 0.44 (-0.06) for AGNs (SFGs), respectively. These measurements are in line with the sample of $10^{9.5}$--$10^{10}M_\odot$, and significant enough to confirm that the AGN feedback triggered in galaxies with dense cores is driving the cold gas depletion in the whole mass range of $10^{9}$--$10^{10}M_\odot$. Future resolved 21~cm and molecular gas surveys targeting low-mass AGN hosts would be a promising way to fully understand the underlying physics of AGN-driven quenching in this mass regime. 

\begin{acknowledgments}
We thank the anonymous reviewer for the helpful comments that improve the presentation of this paper. This work is supported by the National SKA Program of China (grant No. 2020SKA0110100), National Key R\&D Program of China (grant No. 2018YFA0404503), National Science Foundation of China (Nos. 11922305, 11833005, 12073002, 11721303, 12011530159) and the science research grants from the China Manned Space Project with NO. CMS-CSST-2021-A02. We acknowledge the use of the High Performance Computing Resource in the Core Facility for Advanced Research Computing at the Shanghai Astronomical Observatory.
\end{acknowledgments}

\facility{Arecibo, Sloan}

\appendix
\section{Comparison to Literature}\label{sec:comparison}
Our analysis based on $M_{\rm HI}({\rm SFR}|M_\ast)$ reveals strong differences between the SFGs and AGN hosts in the low-mass systems. Some previous works that found no significant differences in the \hi\ content of AGN hosts and non-AGN galaxies selected only galaxies more massive than $10^{10}M_\odot$ \citep{Fabello2011,Gereb2015}, consistent with our findings for these massive galaxies. The non-AGN control galaxies used in previous works \citep{Fabello2011,Ellison2019} include both the SFGs and those galaxies with weak emission lines (low S/N for any of the four emission lines in the BPT diagram). The construction of the control sample and how it is matched to the target sample has the potential to affect the final results. Our sample classification can be more robust without including these low S/N sources in either target or control sample.

The previous literature focus on the comparisons of $f_{\rm HI}(M_\ast)$, or equivalently $\langle M_{\rm HI}(M_\ast)\rangle$ at the same mass bin, which can be in principle obtained from our measurements of $M_{\rm HI}({\rm SFR}|M_\ast)$ as,
\begin{equation}
	\langle M_{\rm HI}(M_\ast)\rangle=\frac{\Sigma N({\rm SFR}|M_\ast)M_{\rm HI}({\rm SFR}|M_\ast)}{\Sigma N({\rm SFR|M_\ast})}, \label{eq:sfr}
\end{equation}
where $N({\rm SFR|M_\ast})$ is the number of galaxies in the SFR bin at the given $M_\ast$ range and the summations are over all the SFR bins. Constructing non-AGN control samples by matching both $M_\ast$ and SFR of the AGN hosts as in the previous work \citep{Ellison2019} is essentially replacing $N({\rm SFR|M_\ast})$ of the non-AGNs in Eq.~(\ref{eq:sfr}) with that of the AGNs. 

In order to compare more directly with these previous works, we apply the \hi-stacking and measure $M_{\rm HI}({\rm SFR}|M_\ast)$ for the non-AGNs by including the SFGs and weak emission line galaxies. Then the stacked $\langle M_{\rm HI}(M_\ast)\rangle$ can be obtained with Eq.~(\ref{eq:sfr}). We show in Table~\ref{tab:sample} the resulting $\langle M_{\rm HI}(M_\ast)\rangle$ for the AGN, SFG, non-AGN, SFG control and non-AGN control samples, where the control samples are using $N({\rm SFR|M_\ast})$ of the AGN hosts in each $M_\ast$ bin. For galaxies in the mass range $10^{9.5}$--$10^{10}M_\odot$, after matching the SFR distributions of the AGN hosts, the non-AGNs have only $0.1$~dex larger $\langle M_{\rm HI}\rangle$ than that of AGNs, despite the large differences between the $M_{\rm HI}({\rm SFR}|M_\ast)$ measurements of SFGs and AGNs. This is due to the fact that the SFR distribution of AGNs peaks at the lower end where the differences between $M_{\rm HI}({\rm SFR}|M_\ast)$ of SFGs and AGNs are much smaller. The effect of including weak emission line galaxies in the non-AGNs is minor for these low-mass galaxies.

The contributions of weak emission line galaxies and the SFR-matching scheme for the more massive samples are slightly different. But the overall effect is the much smaller difference between $\langle M_{\rm HI}(M_\ast)\rangle$ for the AGNs and non-AGN control samples. Our integrated measurements of $\langle M_{\rm HI}(M_\ast)\rangle$ are also consistent with previous work \citep{Ellison2019}. Our experiment here demonstrates that $M_{\rm HI}({\rm SFR}|M_\ast)$ is a much better indicator when comparing the \hi\ masses for different samples than $M_{\rm HI}(M_\ast)$, which would be affected by the distribution of $N({\rm SFR|M_\ast})$.

\begin{table}
	\centering
	\caption{$M_{\rm HI}(M_\ast)$ estimates for the different samples} \label{tab:sample}
	\begin{tabular}{l|ccc}
		\hline
		$\log M_\ast$ & $[9.5,10]$ & $[10,10.5]$ & $[10.5,11]$ \\
		\hline
		$\log M_{\rm HI,SFG}$ & $9.53\pm0.02$ & $9.68\pm0.02$ & $9.90\pm0.04$ \\
		$\log M_{\rm HI,non-AGN}$ & $9.49\pm0.02$ & $9.55\pm0.02$ & $9.66\pm0.04$ \\
		$\log M_{\rm HI,SFG,control}$ & $9.28\pm0.03$ & $9.49\pm0.05$ & $9.64\pm0.06$ \\  
		$\log M_{\rm HI,non-AGN,control}$ & $9.21\pm0.04$ & $9.43\pm0.05$ & $9.67\pm0.06$ \\
		$\log M_{\rm HI,AGN}$ & $9.11\pm0.08$ & $9.35\pm0.04$ & $9.66\pm0.03$ \\
		\hline
	\end{tabular}
	\medskip
	
	The displayed $M_{\rm HI}(M_\ast)$ are obtained through Eq.~\ref{eq:sfr} with the stacked measurements of $M_{\rm HI}({\rm SFR}|M_\ast)$ for different samples. The control samples mean that their SFRs are matched to the corresponding AGNs at the same mass bins. All masses are in units of $M_\odot$.
\end{table}

\section{Robustness of Measurements}
As extensively discussed in previous works \citep{Schawinski2010,Trump2015}, the star formation dilution by the H\,{\sc ii} regions will cause the AGN hosts selected through the BPT diagram systematically biased against low-mass, star-forming and disk-dominated galaxies. The observed low-mass AGN hosts would then be biased towards both higher luminosity and Eddington ratio. As we show in Fig.~\ref{fig:hisfr}, the AGN hosts with larger $L_{\rm [O{\sc III}]}$ and higher Eddington ratios have consistently lower $M_{\rm HI}$ for $M_\ast<10^{10.5}M_\odot$. If we would be able to correct for the selection bias by removing the AGN hosts with low luminosities and Eddington ratios from the observed SFG population, the resulting $M_{\rm HI}$ for the ``pure'' star-forming galaxies would be even higher. It would then lead to larger differences between $M_{\rm HI}$ of the SFGs and high-luminosity AGN hosts for these low-mass galaxies, further supporting our conclusion. We thus emphasize that our conclusion is still robust in light of this potential bias, but the exact $M_{\rm HI}$ offsets between the SFGs and AGN hosts would depend on the appropriate corrections of the selection effect. For more massive galaxies, our results would remain the same, since the AGN selection bias is less severe and there is no trend of $M_{\rm HI}$ with AGN luminosity.     

Since the SDSS fiber aperture size of $3\arcsec$ would cover around 3.7~kpc at the maximum redshift of $z=0.06$, the amount of star formation dilution may be more severe by the relative larger coverage for lower-mass galaxies. We test this effect by limiting our samples to a reduced maximum redshift of $z=0.04$ and find very similar results as in Fig.~\ref{fig:hisfr}. It demonstrates that the aperture effect is not important in this study.

\begin{figure*}
	\centering\includegraphics[width=1\linewidth]{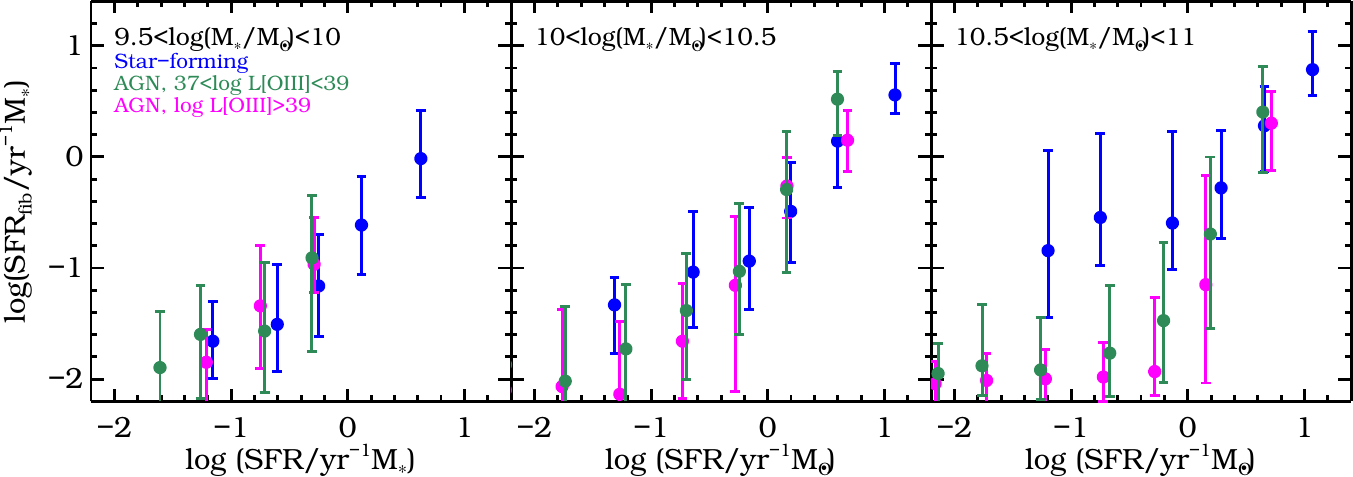}
	\caption{Comparisons of the SED-based total SFRs and those from within the SDSS fibers ($\rm SFR_{fib}$). The comparisons are shown for the SFGs and AGN hosts of different [O{\sc iii}] luminosities as in Fig.~\ref{fig:hisfr}, with the same color coding. We show the median values with the 1$\sigma$ errors in each panel. There is no obvious bias for the $\rm SFR_{fib}$--SFR relations between the SFGs and AGN hosts of $M_\ast<10^{10.5}M_\odot$. For more massive galaxies, the $\rm SFR_{fib}$ estimates for AGN hosts of low SFRs are reaching the lower limit of $\log({\rm SFR_{fib}}/M_\ast)=-11.5$, where the measurements are not reliable.}
	\label{fig:angsfr}
\end{figure*}

Another potential bias for our analysis is that the SFR estimates of the AGN hosts from the GSWLC-2 catalog may suffer from the potential AGN contamination of the emission lines. Although accurate corrections for the contamination is very difficult and uncertain, we can still estimate its influence on our results. For each galaxy in our sample, we further retrieve the SFR estimates within the SDSS fibers (denoted as $\rm SFR_{fib}$) from the MPA-JHU catalog \citep{Brinchmann2004}, which is to check the effect of different SFR estimates, as well as differences between the fiber-based central SFRs and SED-based integrated SFRs. While the $\rm SFR_{fib}$ estimates for the SFGs are based on the emission line fluxes, those for AGN hosts are derived from the D4000 index. 

As the contamination from type-2 AGNs is not important for the D4000-based SFRs, we can use $\rm SFR_{fib}$ of AGNs to check whether there is any systematic bias between SFRs of AGNs and SFGs, despite the large scatters of D4000-based SFRs. If the differences of $M_{\rm HI}$ between SFGs and AGN hosts were caused by the fact that SFRs of AGN hosts are biased high, we would expect to see systematic offsets in the relations between SFR and $\rm SFR_{fib}$ for SFGs and AGN hosts, as well as the systematic trend with the [O{\sc iii}] luminosity. Since the scatter of the $\rm SFR_{fib}$ does not depend on the [O{\sc iii}] luminosity \citep{Rosario2016}, it will not introduce any further bias. 

We show in Figure~\ref{fig:angsfr} comparisons between SFR and $\rm SFR_{fib}$ for SFGs and AGN hosts of different [O{\sc iii}] luminosities. As for all the bins with ${\rm SFR_{fib}}/M_\ast>10^{-11.5}{\rm yr}^{-1}$ (where reliable estimates of ${\rm SFR_{fib}}$ can be obtained), there is not any obvious bias for the $\rm SFR_{fib}$--SFR relations between the SFGs and AGN hosts at different stellar masses, as well as no trend with the [O{\sc iii}] luminosity. We also test that the results are very similar if we limit the redshift range of $0.0025<z<0.04$. We further note that the scatters of $\rm SFR_{fib}$ are very similar for SFGs and AGN hosts, indicating that the D4000-based SFRs for AGNs in our redshift range do not cause any further significant errors with respect to the more accurate $\rm SFR_{fib}$ for SFGs.   

It demonstrates that there is no strong systematic bias in the integral SFRs for the AGN hosts in our current sample. The differences between central SFRs and integral SFRs also do not show any systematic bias towards AGN hosts. Our SFR bins of 0.5~dex are already wide enough to take into account the residual uncertainties of the SFR estimates. Therefore, we conclude that our results of the $M_{\rm HI}$ trend with the AGN luminosity is not likely caused by the systematic bias of SFR estimates for the AGN hosts.

\bibliographystyle{aasjournal}
%\bibliography{ref.bib}

\begin{thebibliography}{}
	\expandafter\ifx\csname natexlab\endcsname\relax\def\natexlab#1{#1}\fi
	\providecommand{\url}[1]{\href{#1}{#1}}
	\providecommand{\dodoi}[1]{doi:~\href{http://doi.org/#1}{\nolinkurl{#1}}}
	\providecommand{\doeprint}[1]{\href{http://ascl.net/#1}{\nolinkurl{http://ascl.net/#1}}}
	\providecommand{\doarXiv}[1]{\href{https://arxiv.org/abs/#1}{\nolinkurl{https://arxiv.org/abs/#1}}}
	
	\bibitem[{{Abazajian} {et~al.}(2009){Abazajian}, {Adelman-McCarthy},
		{Ag{\"u}eros}, {Allam}, {Allende Prieto}, {An}, {Anderson}, {Anderson},
		{Annis}, {Bahcall}, {Bailer-Jones}, {Barentine}, {Bassett}, {Becker},
		{Beers}, {Bell}, {Belokurov}, {Berlind}, {Berman}, {Bernardi}, {Bickerton},
		{Bizyaev}, {Blakeslee}, {Blanton}, {Bochanski}, {Boroski}, {Brewington},
		{Brinchmann}, {Brinkmann}, {Brunner}, {Budav{\'a}ri}, {Carey}, {Carliles},
		{Carr}, {Castander}, {Cinabro}, {Connolly}, {Csabai}, {Cunha}, {Czarapata},
		{Davenport}, {de Haas}, {Dilday}, {Doi}, {Eisenstein}, {Evans}, {Evans},
		{Fan}, {Friedman}, {Frieman}, {Fukugita}, {G{\"a}nsicke}, {Gates},
		{Gillespie}, {Gilmore}, {Gonzalez}, {Gonzalez}, {Grebel}, {Gunn},
		{Gy{\"o}ry}, {Hall}, {Harding}, {Harris}, {Harvanek}, {Hawley}, {Hayes},
		{Heckman}, {Hendry}, {Hennessy}, {Hindsley}, {Hoblitt}, {Hogan}, {Hogg},
		{Holtzman}, {Hyde}, {Ichikawa}, {Ichikawa}, {Im}, {Ivezi{\'c}}, {Jester},
		{Jiang}, {Johnson}, {Jorgensen}, {Juri{\'c}}, {Kent}, {Kessler}, {Kleinman},
		{Knapp}, {Konishi}, {Kron}, {Krzesinski}, {Kuropatkin}, {Lampeitl},
		{Lebedeva}, {Lee}, {Lee}, {French Leger}, {L{\'e}pine}, {Li}, {Lima}, {Lin},
		{Long}, {Loomis}, {Loveday}, {Lupton}, {Magnier}, {Malanushenko},
		{Malanushenko}, {Mandelbaum}, {Margon}, {Marriner}, {Mart{\'\i}nez-Delgado},
		{Matsubara}, {McGehee}, {McKay}, {Meiksin}, {Morrison}, {Mullally}, {Munn},
		{Murphy}, {Nash}, {Nebot}, {Neilsen}, {Newberg}, {Newman}, {Nichol},
		{Nicinski}, {Nieto-Santisteban}, {Nitta}, {Okamura}, {Oravetz}, {Ostriker},
		{Owen}, {Padmanabhan}, {Pan}, {Park}, {Pauls}, {Peoples}, {Percival}, {Pier},
		{Pope}, {Pourbaix}, {Price}, {Purger}, {Quinn}, {Raddick}, {Re Fiorentin},
		{Richards}, {Richmond}, {Riess}, {Rix}, {Rockosi}, {Sako}, {Schlegel},
		{Schneider}, {Scholz}, {Schreiber}, {Schwope}, {Seljak}, {Sesar}, {Sheldon},
		{Shimasaku}, {Sibley}, {Simmons}, {Sivarani}, {Allyn Smith}, {Smith},
		{Smol{\v{c}}i{\'c}}, {Snedden}, {Stebbins}, {Steinmetz}, {Stoughton},
		{Strauss}, {SubbaRao}, {Suto}, {Szalay}, {Szapudi}, {Szkody}, {Tanaka},
		{Tegmark}, {Teodoro}, {Thakar}, {Tremonti}, {Tucker}, {Uomoto}, {Vanden
			Berk}, {Vandenberg}, {Vidrih}, {Vogeley}, {Voges}, {Vogt}, {Wadadekar},
		{Watters}, {Weinberg}, {West}, {White}, {Wilhite}, {Wonders}, {Yanny},
		{Yocum}, {York}, {Zehavi}, {Zibetti}, \& {Zucker}}]{SDSSDR7}
	{Abazajian}, K.~N., {Adelman-McCarthy}, J.~K., {Ag{\"u}eros}, M.~A., {et~al.}
	2009, \apjs, 182, 543, \dodoi{10.1088/0067-0049/182/2/543}
	
	\bibitem[{{Baldwin} {et~al.}(1981){Baldwin}, {Phillips}, \&
		{Terlevich}}]{Baldwin1981}
	{Baldwin}, J.~A., {Phillips}, M.~M., \& {Terlevich}, R. 1981, \pasp, 93, 5,
	\dodoi{10.1086/130766}
	
	\bibitem[{{Barro} {et~al.}(2013){Barro}, {Faber}, {P{\'e}rez-Gonz{\'a}lez},
		{Koo}, {Williams}, {Kocevski}, {Trump}, {Mozena}, {McGrath}, {van der Wel},
		{Wuyts}, {Bell}, {Croton}, {Ceverino}, {Dekel}, {Ashby}, {Cheung},
		{Ferguson}, {Fontana}, {Fang}, {Giavalisco}, {Grogin}, {Guo}, {Hathi},
		{Hopkins}, {Huang}, {Koekemoer}, {Kartaltepe}, {Lee}, {Newman}, {Porter},
		{Primack}, {Ryan}, {Rosario}, {Somerville}, {Salvato}, \& {Hsu}}]{Barro2013}
	{Barro}, G., {Faber}, S.~M., {P{\'e}rez-Gonz{\'a}lez}, P.~G., {et~al.} 2013,
	\apj, 765, 104, \dodoi{10.1088/0004-637X/765/2/104}
	
	\bibitem[{{Bradford} {et~al.}(2018){Bradford}, {Geha}, {Greene}, {Reines}, \&
		{Dickey}}]{Bradford2018}
	{Bradford}, J.~D., {Geha}, M.~C., {Greene}, J.~E., {Reines}, A.~E., \&
	{Dickey}, C.~M. 2018, \apj, 861, 50, \dodoi{10.3847/1538-4357/aac88d}
	
	\bibitem[{{Brinchmann} {et~al.}(2004){Brinchmann}, {Charlot}, {White},
		{Tremonti}, {Kauffmann}, {Heckman}, \& {Brinkmann}}]{Brinchmann2004}
	{Brinchmann}, J., {Charlot}, S., {White}, S.~D.~M., {et~al.} 2004, \mnras, 351,
	1151, \dodoi{10.1111/j.1365-2966.2004.07881.x}
	
	\bibitem[{{Brown} {et~al.}(2017){Brown}, {Catinella}, {Cortese}, {Lagos},
		{Dav{\'e}}, {Kilborn}, {Haynes}, {Giovanelli}, \&
		{Rafieferantsoa}}]{Brown2017}
	{Brown}, T., {Catinella}, B., {Cortese}, L., {et~al.} 2017, \mnras, 466, 1275,
	\dodoi{10.1093/mnras/stw2991}
	
	\bibitem[{{Charlot} \& {Fall}(2000)}]{Charlot2000}
	{Charlot}, S., \& {Fall}, S.~M. 2000, \apj, 539, 718, \dodoi{10.1086/309250}
	
	\bibitem[{{Chen} {et~al.}(2020){Chen}, {Faber}, {Koo}, {Somerville}, {Primack},
		{Dekel}, {Rodr{\'\i}guez-Puebla}, {Guo}, {Barro}, {Kocevski}, {van der Wel},
		{Woo}, {Bell}, {Fang}, {Ferguson}, {Giavalisco}, {Huertas-Company}, {Jiang},
		{Kassin}, {Lin}, {Liu}, {Luo}, {Luo}, {Pacifici}, {Pandya}, {Salim}, {Shu},
		{Tacchella}, {Terrazas}, \& {Yesuf}}]{Chen2020}
	{Chen}, Z., {Faber}, S.~M., {Koo}, D.~C., {et~al.} 2020, \apj, 897, 102,
	\dodoi{10.3847/1538-4357/ab9633}
	
	\bibitem[{{Cui} {et~al.}(2021){Cui}, {Dav{\'e}}, {Peacock},
		{Angl{\'e}s-Alc{\'a}zar}, \& {Yang}}]{Cui2021}
	{Cui}, W., {Dav{\'e}}, R., {Peacock}, J.~A., {Angl{\'e}s-Alc{\'a}zar}, D., \&
	{Yang}, X. 2021, Nature Astronomy, 5, 1078,
	\dodoi{10.1038/s41550-021-01456-3}
	
	\bibitem[{{Dekel} \& {Burkert}(2014)}]{Dekel2014}
	{Dekel}, A., \& {Burkert}, A. 2014, \mnras, 438, 1870,
	\dodoi{10.1093/mnras/stt2331}
	
	\bibitem[{{Ellison} {et~al.}(2019){Ellison}, {Brown}, {Catinella}, \&
		{Cortese}}]{Ellison2019}
	{Ellison}, S.~L., {Brown}, T., {Catinella}, B., \& {Cortese}, L. 2019, \mnras,
	482, 5694, \dodoi{10.1093/mnras/sty3139}
	
	\bibitem[{{Ellison} {et~al.}(2021){Ellison}, {Wong}, {S{\'a}nchez}, {Colombo},
		{Bolatto}, {Barrera-Ballesteros}, {Garc{\'\i}a-Benito}, {Kalinova}, {Luo},
		{Rubio}, \& {Vogel}}]{Ellison2021}
	{Ellison}, S.~L., {Wong}, T., {S{\'a}nchez}, S.~F., {et~al.} 2021, \mnras, 505,
	L46, \dodoi{10.1093/mnrasl/slab047}
	
	\bibitem[{{Fabello} {et~al.}(2011){Fabello}, {Kauffmann}, {Catinella},
		{Giovanelli}, {Haynes}, {Heckman}, \& {Schiminovich}}]{Fabello2011}
	{Fabello}, S., {Kauffmann}, G., {Catinella}, B., {et~al.} 2011, \mnras, 416,
	1739, \dodoi{10.1111/j.1365-2966.2011.18825.x}
	
	\bibitem[{{Fabian}(2012)}]{Fabian2012}
	{Fabian}, A.~C. 2012, \araa, 50, 455,
	\dodoi{10.1146/annurev-astro-081811-125521}
	
	\bibitem[{{F{\"o}rster Schreiber} \& {Wuyts}(2020)}]{Schreiber2020}
	{F{\"o}rster Schreiber}, N.~M., \& {Wuyts}, S. 2020, \araa, 58, 661,
	\dodoi{10.1146/annurev-astro-032620-021910}
	
	\bibitem[{{Ger{\'e}b} {et~al.}(2015){Ger{\'e}b}, {Morganti}, {Oosterloo},
		{Hoppmann}, \& {Staveley-Smith}}]{Gereb2015}
	{Ger{\'e}b}, K., {Morganti}, R., {Oosterloo}, T.~A., {Hoppmann}, L., \&
	{Staveley-Smith}, L. 2015, \aap, 580, A43,
	\dodoi{10.1051/0004-6361/201424810}
	
	\bibitem[{{Giovanelli} {et~al.}(2005){Giovanelli}, {Haynes}, {Kent},
		{Perillat}, {Saintonge}, {Brosch}, {Catinella}, {Hoffman}, {Stierwalt},
		{Spekkens}, {Lerner}, {Masters}, {Momjian}, {Rosenberg}, {Springob},
		{Boselli}, {Charmandaris}, {Darling}, {Davies}, {Garcia Lambas}, {Gavazzi},
		{Giovanardi}, {Hardy}, {Hunt}, {Iovino}, {Karachentsev}, {Karachentseva},
		{Koopmann}, {Marinoni}, {Minchin}, {Muller}, {Putman}, {Pantoja}, {Salzer},
		{Scodeggio}, {Skillman}, {Solanes}, {Valotto}, {van Driel}, \& {van
			Zee}}]{Giovanelli2005}
	{Giovanelli}, R., {Haynes}, M.~P., {Kent}, B.~R., {et~al.} 2005, \aj, 130,
	2598, \dodoi{10.1086/497431}
	
	\bibitem[{{Guo} {et~al.}(2020){Guo}, {Jones}, {Haynes}, \& {Fu}}]{Guo2020}
	{Guo}, H., {Jones}, M.~G., {Haynes}, M.~P., \& {Fu}, J. 2020, \apj, 894, 92,
	\dodoi{10.3847/1538-4357/ab886f}
	
	\bibitem[{{Guo} {et~al.}(2021){Guo}, {Jones}, {Wang}, \& {Lin}}]{Guo2021}
	{Guo}, H., {Jones}, M.~G., {Wang}, J., \& {Lin}, L. 2021, \apj, 918, 53,
	\dodoi{10.3847/1538-4357/ac062e}
	
	\bibitem[{{Harrison}(2017)}]{Harrison2017}
	{Harrison}, C.~M. 2017, Nature Astronomy, 1, 0165,
	\dodoi{10.1038/s41550-017-0165}
	
	\bibitem[{{Haynes} {et~al.}(2018){Haynes}, {Giovanelli}, {Kent}, {Adams},
		{Balonek}, {Craig}, {Fertig}, {Finn}, {Giovanardi}, {Hallenbeck}, {Hess},
		{Hoffman}, {Huang}, {Jones}, {Koopmann}, {Kornreich}, {Leisman}, {Miller},
		{Moorman}, {O'Connor}, {O'Donoghue}, {Papastergis}, {Troischt}, {Stark}, \&
		{Xiao}}]{Haynes2018}
	{Haynes}, M.~P., {Giovanelli}, R., {Kent}, B.~R., {et~al.} 2018, \apj, 861, 49,
	\dodoi{10.3847/1538-4357/aac956}
	
	\bibitem[{{Heckman} \& {Best}(2014)}]{Heckman2014}
	{Heckman}, T.~M., \& {Best}, P.~N. 2014, \araa, 52, 589,
	\dodoi{10.1146/annurev-astro-081913-035722}
	
	\bibitem[{{Heckman} {et~al.}(2004){Heckman}, {Kauffmann}, {Brinchmann},
		{Charlot}, {Tremonti}, \& {White}}]{Heckman2004}
	{Heckman}, T.~M., {Kauffmann}, G., {Brinchmann}, J., {et~al.} 2004, \apj, 613,
	109, \dodoi{10.1086/422872}
	
	\bibitem[{{Hickox} {et~al.}(2014){Hickox}, {Mullaney}, {Alexander}, {Chen},
		{Civano}, {Goulding}, \& {Hainline}}]{Hickox2014}
	{Hickox}, R.~C., {Mullaney}, J.~R., {Alexander}, D.~M., {et~al.} 2014, \apj,
	782, 9, \dodoi{10.1088/0004-637X/782/1/9}
	
	\bibitem[{{Jaff{\'e}} {et~al.}(2015){Jaff{\'e}}, {Smith}, {Candlish},
		{Poggianti}, {Sheen}, \& {Verheijen}}]{Jaffe2015}
	{Jaff{\'e}}, Y.~L., {Smith}, R., {Candlish}, G.~N., {et~al.} 2015, \mnras, 448,
	1715, \dodoi{10.1093/mnras/stv100}
	
	\bibitem[{{Karouzos} {et~al.}(2016){Karouzos}, {Woo}, \& {Bae}}]{Karouzos2016}
	{Karouzos}, M., {Woo}, J.-H., \& {Bae}, H.-J. 2016, \apj, 819, 148,
	\dodoi{10.3847/0004-637X/819/2/148}
	
	\bibitem[{{Kauffmann} \& {Heckman}(2009)}]{Kauffmann2009}
	{Kauffmann}, G., \& {Heckman}, T.~M. 2009, \mnras, 397, 135,
	\dodoi{10.1111/j.1365-2966.2009.14960.x}
	
	\bibitem[{{Kauffmann} {et~al.}(2003){Kauffmann}, {Heckman}, {Tremonti},
		{Brinchmann}, {Charlot}, {White}, {Ridgway}, {Brinkmann}, {Fukugita}, {Hall},
		{Ivezi{\'c}}, {Richards}, \& {Schneider}}]{Kauffmann2003}
	{Kauffmann}, G., {Heckman}, T.~M., {Tremonti}, C., {et~al.} 2003, \mnras, 346,
	1055, \dodoi{10.1111/j.1365-2966.2003.07154.x}
	
	\bibitem[{{Lim} {et~al.}(2017){Lim}, {Mo}, {Lu}, {Wang}, \& {Yang}}]{Lim2017}
	{Lim}, S.~H., {Mo}, H.~J., {Lu}, Y., {Wang}, H., \& {Yang}, X. 2017, \mnras,
	470, 2982, \dodoi{10.1093/mnras/stx1462}
	
	\bibitem[{{Novak} {et~al.}(2011){Novak}, {Ostriker}, \& {Ciotti}}]{Novak2011}
	{Novak}, G.~S., {Ostriker}, J.~P., \& {Ciotti}, L. 2011, \apj, 737, 26,
	\dodoi{10.1088/0004-637X/737/1/26}
	
	\bibitem[{{Rosario} {et~al.}(2016){Rosario}, {Mendel}, {Ellison}, {Lutz}, \&
		{Trump}}]{Rosario2016}
	{Rosario}, D.~J., {Mendel}, J.~T., {Ellison}, S.~L., {Lutz}, D., \& {Trump},
	J.~R. 2016, \mnras, 457, 2703, \dodoi{10.1093/mnras/stw096}
	
	\bibitem[{{Salim} {et~al.}(2018){Salim}, {Boquien}, \& {Lee}}]{Salim2018}
	{Salim}, S., {Boquien}, M., \& {Lee}, J.~C. 2018, \apj, 859, 11,
	\dodoi{10.3847/1538-4357/aabf3c}
	
	\bibitem[{{Salim} {et~al.}(2016){Salim}, {Lee}, {Janowiecki}, {da Cunha},
		{Dickinson}, {Boquien}, {Burgarella}, {Salzer}, \& {Charlot}}]{Salim2016}
	{Salim}, S., {Lee}, J.~C., {Janowiecki}, S., {et~al.} 2016, \apjs, 227, 2,
	\dodoi{10.3847/0067-0049/227/1/2}
	
	\bibitem[{{Schawinski} {et~al.}(2010){Schawinski}, {Urry}, {Virani}, {Coppi},
		{Bamford}, {Treister}, {Lintott}, {Sarzi}, {Keel}, {Kaviraj}, {Cardamone},
		{Masters}, {Ross}, {Andreescu}, {Murray}, {Nichol}, {Raddick}, {Slosar},
		{Szalay}, {Thomas}, \& {Vandenberg}}]{Schawinski2010}
	{Schawinski}, K., {Urry}, C.~M., {Virani}, S., {et~al.} 2010, \apj, 711, 284,
	\dodoi{10.1088/0004-637X/711/1/284}
	
	\bibitem[{{Shangguan} {et~al.}(2020){Shangguan}, {Ho}, {Bauer}, {Wang}, \&
		{Treister}}]{Shangguan2020}
	{Shangguan}, J., {Ho}, L.~C., {Bauer}, F.~E., {Wang}, R., \& {Treister}, E.
	2020, \apj, 899, 112, \dodoi{10.3847/1538-4357/aba8a1}
	
	\bibitem[{{Shi} {et~al.}(2021){Shi}, {Li}, {Yuan}, \& {Zhu}}]{Shi2021}
	{Shi}, F., {Li}, Z., {Yuan}, F., \& {Zhu}, B. 2021, Nature Astronomy, 5, 928,
	\dodoi{10.1038/s41550-021-01394-0}
	
	\bibitem[{{Stevens} {et~al.}(2019){Stevens}, {Diemer}, {Lagos}, {Nelson},
		{Pillepich}, {Brown}, {Catinella}, {Hernquist}, {Weinberger}, {Vogelsberger},
		\& {Marinacci}}]{Stevens2019}
	{Stevens}, A. R.~H., {Diemer}, B., {Lagos}, C. d.~P., {et~al.} 2019, \mnras,
	483, 5334, \dodoi{10.1093/mnras/sty3451}
	
	\bibitem[{{Tal} {et~al.}(2014){Tal}, {Dekel}, {Oesch}, {Muzzin}, {Brammer},
		{van Dokkum}, {Franx}, {Illingworth}, {Leja}, {Magee}, {Marchesini},
		{Momcheva}, {Nelson}, {Patel}, {Quadri}, {Rix}, {Skelton}, {Wake}, \&
		{Whitaker}}]{Tal2014}
	{Tal}, T., {Dekel}, A., {Oesch}, P., {et~al.} 2014, \apj, 789, 164,
	\dodoi{10.1088/0004-637X/789/2/164}
	
	\bibitem[{{Tremaine} {et~al.}(2002){Tremaine}, {Gebhardt}, {Bender}, {Bower},
		{Dressler}, {Faber}, {Filippenko}, {Green}, {Grillmair}, {Ho}, {Kormendy},
		{Lauer}, {Magorrian}, {Pinkney}, \& {Richstone}}]{Tremaine2002}
	{Tremaine}, S., {Gebhardt}, K., {Bender}, R., {et~al.} 2002, \apj, 574, 740,
	\dodoi{10.1086/341002}
	
	\bibitem[{{Trump} {et~al.}(2015){Trump}, {Sun}, {Zeimann}, {Luck}, {Bridge},
		{Grier}, {Hagen}, {Juneau}, {Montero-Dorta}, {Rosario}, {Brandt},
		{Ciardullo}, \& {Schneider}}]{Trump2015}
	{Trump}, J.~R., {Sun}, M., {Zeimann}, G.~R., {et~al.} 2015, \apj, 811, 26,
	\dodoi{10.1088/0004-637X/811/1/26}
	
	\bibitem[{{Veilleux} \& {Osterbrock}(1987)}]{Veilleux1987}
	{Veilleux}, S., \& {Osterbrock}, D.~E. 1987, \apjs, 63, 295,
	\dodoi{10.1086/191166}
	
	\bibitem[{{Wang} {et~al.}(2016){Wang}, {Koribalski}, {Serra}, {van der Hulst},
		{Roychowdhury}, {Kamphuis}, \& {Chengalur}}]{Wang2016}
	{Wang}, J., {Koribalski}, B.~S., {Serra}, P., {et~al.} 2016, \mnras, 460, 2143,
	\dodoi{10.1093/mnras/stw1099}
	
	\bibitem[{{Weinberger} {et~al.}(2018){Weinberger}, {Springel}, {Pakmor},
		{Nelson}, {Genel}, {Pillepich}, {Vogelsberger}, {Marinacci}, {Naiman},
		{Torrey}, \& {Hernquist}}]{Weinberger2018}
	{Weinberger}, R., {Springel}, V., {Pakmor}, R., {et~al.} 2018, \mnras, 479,
	4056, \dodoi{10.1093/mnras/sty1733}
	
	\bibitem[{{Wetzel} {et~al.}(2013){Wetzel}, {Tinker}, {Conroy}, \& {van den
			Bosch}}]{Wetzel2013}
	{Wetzel}, A.~R., {Tinker}, J.~L., {Conroy}, C., \& {van den Bosch}, F.~C. 2013,
	\mnras, 432, 336, \dodoi{10.1093/mnras/stt469}
	
	\bibitem[{{Zolotov} {et~al.}(2015){Zolotov}, {Dekel}, {Mandelker}, {Tweed},
		{Inoue}, {DeGraf}, {Ceverino}, {Primack}, {Barro}, \& {Faber}}]{Zolotov2015}
	{Zolotov}, A., {Dekel}, A., {Mandelker}, N., {et~al.} 2015, \mnras, 450, 2327,
	\dodoi{10.1093/mnras/stv740}
	
\end{thebibliography}

\end{document}